\documentclass[sigconf,natbib=false]{acmart}

\usepackage{graphicx}
\usepackage{pgfplots}  
\usepackage{xcolor,colortbl}
\usepackage{url}
\usepackage{parskip}

\usepackage{enumitem}
\usepackage{graphicx}
\usepackage{verbatim}
\usepackage{adjustbox}

\usepackage[noend]{algorithmic}
\usepackage{algorithm}
\usepackage{amsthm}
\usepackage{lipsum}
\usepackage{parskip}

\usepackage[style=ACM-Reference-Format,backend=bibtex,sorting=none]{biblatex}
\pgfplotsset{width=5.5cm,compat=1.7}  
\setcopyright{none} 

\setcopyright{none}
\renewcommand\footnotetextcopyrightpermission[1]{} 
\begin{document}

\title{TrustToken, a Trusted SoC solution for Non-Trusted Intellectual Property (IP)s }


\author{Muhammed kawser Ahmed}
\authornote{}
\email{muhammed.kawsera@ufl.edu}
\affiliation{%
  \institution{Computer Engineering, University of Florida}
  \streetaddress{}
  \city{Gainesville}
  \state{FL}
  \country{USA}
  \postcode{32603}
}

\author{Sujan Kumar Saha}
\authornote{}
\email{sujansaha@ufl.edu}
\affiliation{%
 \institution{Computer Engineering, University of Florida}
  \streetaddress{}
 \city{Gainesville}
 \state{FL}
  \country{USA}
  \postcode{32603}
}

\author{Dr. Christophe Bobda }
\email{cbobda@ufl.edu}
\affiliation{%
   \institution{Computer Engineering, University of Florida}
 \streetaddress{}
 \city{Gainesville}
 \state{FL}
 \country{USA}
 \postcode{32603}
}


\begin{abstract}
 Secure and trustworthy execution in heterogeneous SoCs is a major priority in the modern computing system. Security of SoCs mainly addresses two broad layers of trust issues: 1. Protection against hardware security threats(Side-channel, IP Privacy, Cloning, Fault Injection, and Denial of Service); and 2. Protection against malicious software attacks running on SoC processors.  To resist malicious software-level attackers from gaining unauthorized access and compromising security, we propose a root of trust-based trusted execution mechanism \textbf{\textit{(named as \textbf{TrustToken}) }}. TrustToken builds a security block to provide a root of trust-based IP security: secure key generation and truly random source.
 \textbf{TrustToken} only allows trusted
communication between the non-trusted third-party IP and the rest of the SoC world by providing essential security features, i.e., secure, isolated execution, and trusted user interaction. The proposed design achieves this by interconnecting the third-party IP interface to \textbf{TrustToken} Controller and checking IP authorization(Token) signals \texttt{`correctness'} at run-time.  \textbf{TrustToken} architecture shows a very low overhead resource utilization LUT (618,  1.16 \%), FF (44, 0.04 \%), and BUFG (2 , 6.25\%) in implementation. The experiment results show that TrustToken can provide a secure, low-cost, and trusted solution for non-trusted SoC IPs.

\end{abstract}

\settopmatter{printfolios=true}
	\maketitle
\pagestyle{plain}


\section{Introduction}

Often non-trusted third-party IP cores or EDA tools are integrated into different stages of SoC development life and are susceptible to numerous attacks, such as HT injection, IP piracy, cloning, tampering and unauthorized access \cite{trojan_1}, 
To prevent malicious attackers from gaining unauthorized access and leaking sensitive information, modern SoCs are equipped with sandboxing mechanism where applications and OS are executed in a isolated trusted environment \cite{trustzone_white}. ARM TrustZone is the industry leading sandboxing mechanism which is widely available in mobile and heterogenous SoC devices for providing trusted execution environment (TEE) where untrusted IP core is executed in a isolated secure processor along with separated Memory, Cache and Bus system. Leveraging ARM TrustZone technology, TEE is extended for security measures in many academic projects and industrial applications such as Samsung Knox \cite{samsung_knox}, Android's Keystore \cite{android_tee} , OP-TEE \cite{op_tee}, Xilinx TrustZone \cite{xilinx_trustzone} etc.

A SoC CPU which includes any Trusted Execution Environment (TEE) technology e.g. ARM TrustZone, Knox, Xilinx TrustZone etc. don't provide any root of trust based secure mechanism, where any running software is verified and trustworthy. Instead of root of trust based authorization existing TEE technologies focuses on only isolating the environment by partitioning the CPU, memory and system bus. Some of the current TrustZone based technologies assumes the availability of a secure storage devices for storing secret keys which can only be accessed by the secure world entity and serve as the root of the trust. On many modern SoC and mobile devices unfortunately that secured storage devices is not available. For establishing root of trust based trusted environment, the device key should be stored securely and available after a reboot. In many reconfigurable SoCs e.g. Xilinx Zynq - 7000 SoC \cite{trustzone_white}, UltraScale+ 
FreeScale etc., the secure key is stored either in battery backed RAM (BBRAM) or eFuse medium. However, there are some bottlenecks in this method and not recommended for following disadvantages: 1) These mechanism still need to provide secure random key generation by random number generator(RNG) which will serve as a root of trust. 2) eFuse is a non-changeble memory where the key cannot be updated. 3) BBRAM methods needs a physical battery to be placed for storage.

\begin{figure}[h]
	\centerline{\includegraphics[width=7cm]{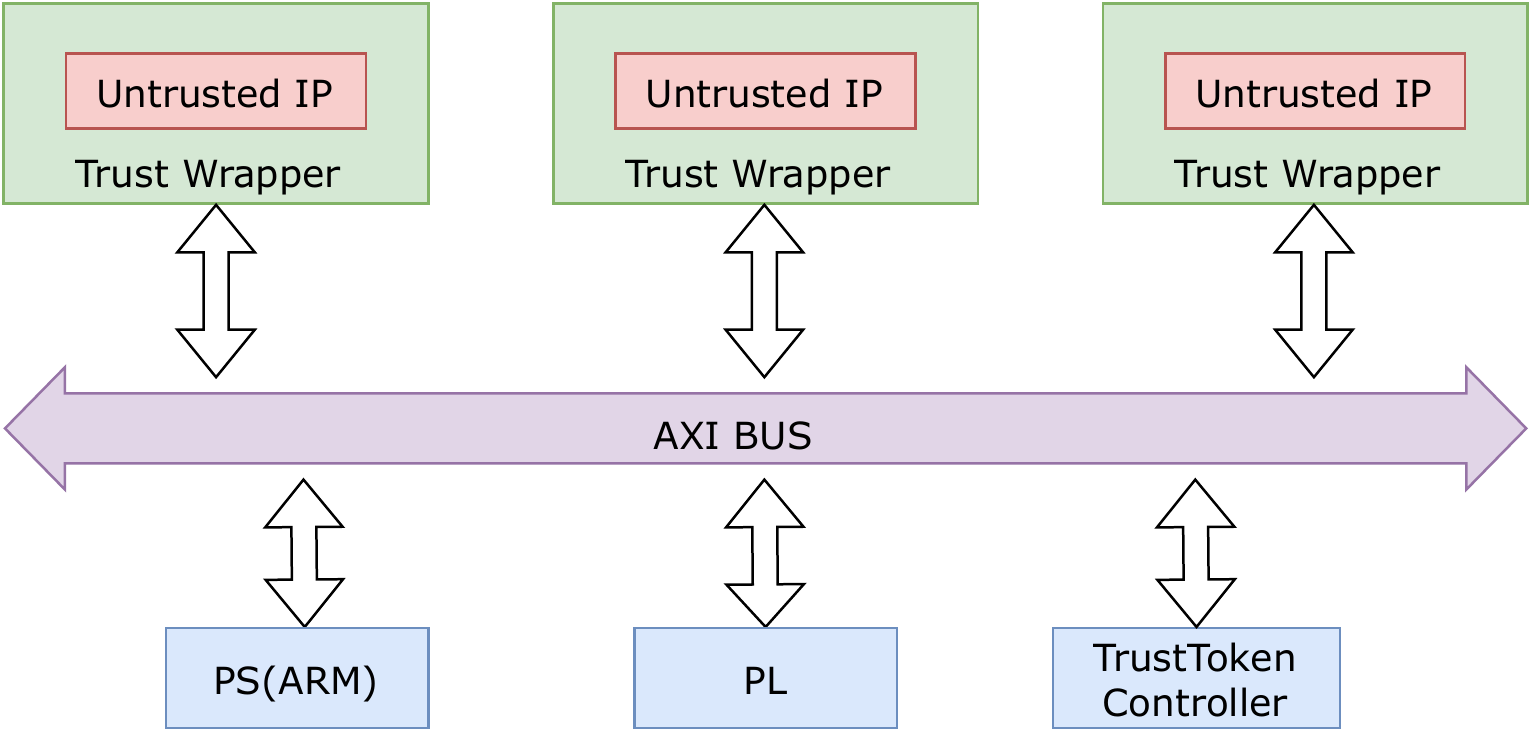}}

	\caption{\textbf{Proposed TrustToken architecture} }
	\label{trust_wrapper}
\end{figure}

In this paper, we propose a novel and efficient trusted technology \textbf{\textbf{TrustToken}} to overcome above disadvantages. We used an asymmetric cryptographic solution that can generate identification tokens in runtime conditions without using any non-volatile memory or any secure boot system.  We are inspired and motivated by Google Chromium sand-boxing \cite{google_sandbox} concept to establish a secure execution environment in a SoC background by assigning Tokens for each IP core. \textbf{\textbf{TrustToken}} is a security mechanism that allows to execute a non trusted third party IP in a closed and monitored environment. If a malicious attacker is able to exploit the access control of the IP in a way that lets him run arbitrary alter on the IP design, the \textbf{TrustToken} would help prevent this incident from causing damage to the system. This is achieved by wrapping a non-trusted IP with a security wrapper shown in Fig. \ref{trust_wrapper}. This security wrapper is connected with \textbf{TrustToken} Controller that performs the security evaluation of each connected non-trusted IP core 
Token IDs are randomly generated by exploiting hardware process variation and assigned to every IP connected by the \textbf{TrustToken} controller in the boot stage. 
\textbf{TrustToken} controller also incorporates a Physical Unclonable Function (PUF) IP block to generate root of trust based ubiquitous token keys. Each IP core has its own token, and the \textbf{TrustToken} controller compares this token to deny or grant access to the rest of the SoC system. This token access signal acts like a security ID card for the untrusted IP core and must provide in each data transaction access request.


\section{Related Work }

In academic research, the first isolation mechanism was proposed by Huffmire in the paper \cite{physicalisolation_huffmire}. In this proposed security mechanism named "Moats and drawbridges," a reconfigurable SoC is configured by creating a fence around the IP with a block of wires (moats). This fenced region can only communicate with another region via a "drawbridge" tunnel. Hategekimana et al.  \cite{isolation_1} proposed the integration of nontrusted IPs in systems within a hardware sandbox to prevent malicious attacks by HTs. But the weakness of sandbox-based security is, it only allows allowable interactions with pre-defined access control definitions, and hence only specific violations are restricted.  

  Zhao et al. proposed a prototype that extends the ARM TrustZone technology to establish a root of trust-based authorization (secure storage, randomness, and secure boot) by using Mobile SRAM PUFs \cite{Zhao_trustzone_sram}. One of the disadvantages of this method is that SRAM is a poor, unreliable PUF solution and needs additional error correction code (ECC) algorithms to be employed with a helper data mechanism which increases the post-processing cost. In the paper \cite{Zhao_trustzone_token}, Zhao et al. extended their previous work and proposed a two-factor authentication protocol in mobile SoC platform by integrating separate Hash, RSA, and AES modules. The implementation latency on this work was poor and could not fit the real-world SoC IP security measures. In work \cite{basak_2017}, Abhishek et al. proposed a security wrapper-based framework around a third-party IP core, and within the security wrapper, 
 security policies were added to prevent malicious activities in
the system. This work focuses mainly on verification and debug purposes and does not fit for runtime trojan mitigation or software level attacks preventions due to high overhead and latency. In the proposed work \cite{ray_policy_2015}, authors mentioned a secured architecture of multiple IPs integrated with the OS kernel and applications, which lacks proper implementation. 

The major drawback of the ARM TrustZone architecture is sharing same peripherals such as Ethernet, DRAM, UART, etc., which are susceptible to row-hammer and Denial-of-service attacks \cite{pinto_arm_2019}.
The primary security concern of ARM TrustZone technology is its weak and inefficient authentication mechanism. 
Several research works have published, indicating unauthorized gain of  kernel-level privilege in ARM TrustZone platforms from normal world environment \cite{pinto_arm_2019}.

\section{Background}
\subsection{Physical Unclonable Function}
Physical Unclonable Function exploits the manufacturing variation of a silicon nanocircuit to generate unique and ubiquitous keys \cite{kawser_puf}. PUF can be used for cryptographic operations such as authentication, random number generation, authorization, etc. The idea behind the PUF is that one (or more) device that is identical by design will have different electrical characteristics due to manufacturing variation.

To evaluate PUF generated keys performance, the three most common metrics are used. They are Randomness, Bit Error Rate, and Uniqueness. 
A strong PUF design has many challenge-response pairs (CRPs)  generated from a single device, and normally weak PUFs support a relatively small number of CRPs. Compared to other crypto measures such as AES, SHA, MD5, or HASH functions, PUFs exploit limited hardware resources (LUTs, GATES).   
\vspace{-2.5 mm }
\subsection{ARM TRUSTZONE}
ARM TrustZone technology refers to a secure execution
environment (SEE) \cite{trustzone_white} where an environment is provided to
isolate both trusted and non trusted software and hardware. It
is also referred to as Trusted Execution Environment (TEE), and
it has a monitor that controls the interactions between these
two separate worlds. TEE TrustZone uses two physically separate processors
dedicated to the trusted and non-trusted world in an embedded security system. The
major drawback of this architecture is they share the same
peripherals such as Ethernet, DRAM, UART, etc. ARM TrustZone is a combination
of some IP blocks which allows a partition between sets of I/O
Peripherals, Processors, and Memory into two different worlds.
In ARM TrustZone platform, two NS bit registers is dedicated to implement the isolation of a software process. \cite{trustzone_white}.

\subsection{Hardware Trojan and Design For Trust}

A hardware Trojan (HT) is stated as a malicious attacker who has intentionally modified a system circuit and cause alteration from expected behavior while the circuit is implemented \cite{trojan_2}. HT poses a severe threat to SoC design as it can leak sensitive information or change the specification of a circuit in run time conditions. HT can create a emergency situation by degrading the overall system functionality of the circuit. Often this HT is deployed in stealth mode and activated in rare conditions, making it very difficult to patrol its harmful effects in the verification stage. Many researchers have come forward to classifying hardware Trojans and their structure based on their characteristics. One of the best classifications is according to the activation (referred to as Trojan Trigger), and payload mechanism ground \cite{trojan_1}.

\section{Threat Model and System Properties}

\label{sec:threat}
\noindent

Before digging into the proposed architecture, we considered some threat models. Our threat model can be divided into two different explicit scenarios: Hardware Trojan and Illegal software access.  
In considering the probable first threat model, we consider every IP as non-trusted and capable of inserting hidden malicious Trojan components inside the IP component. They can act in rare conditions. We assume that they are only activated in run-time environment situations and executed covertly from the interior design of the IP. We assume that the SoC IP integrator division is trusted. All other entities in the supply chain process, such as designers, foundries, manufacturers, and validators, can insert this Trojan at any level of the IC life cycle. In this scenario, \textbf{TrustToken} can provides protection against unauthorized data access, access control, modifications, and leaking of any sensitive information from the IP core to the outsider world.

\begin{figure}[h]
	\includegraphics[width=6cm]{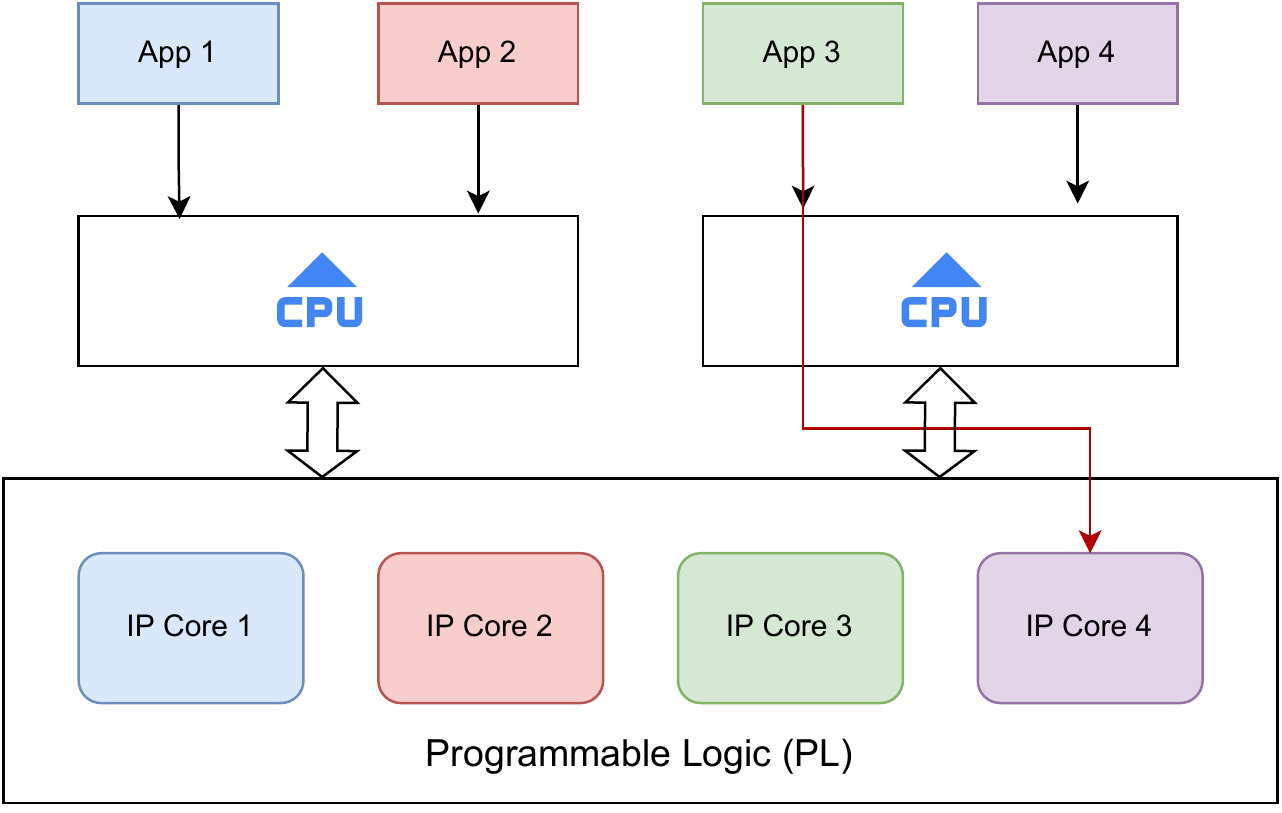}
	\caption{Illegal software access request by Application 3 running on CPU}
	\label{fig:illegal_access}
\end{figure}

In the second scenario, we assume a malicious attacker, who can gain illegal software access from the embedded SoC world, and leaks/modifies/steals sensitive information. For example, figure \ref{fig:illegal_access} shows an example scenario of malicious unauthorized access. In this figure, four software-level applications are running on two separate CPUs placed in the same SoC system. Four custom IPs are added in the hardware level design, which can be accessed from the software side and marked as the same color as the corresponding application.  Access request of IP core 4(four) by software Application 3 (three) can be flagged as illegal and can be isolated by the proposed architecture model.   

In this article, we did not consider the physical attacks performed by physical equipment, which are out of the scope of this article. The attack scenarios did not cover attacks related to hardware components like side-channel attacks, probing attacks, snooping, timing attacks, denial-of-service, fault injection, etc. In summary, describing our architecture, we have taken these threat models into account : 

\begin{enumerate}[leftmargin=*]

	\item Any malicious HT hiding inside of an IP core, trying to execute in runtime environments. We assume that, hidden HT can bypass the existing CAD tools and can be undetected until payload condition is triggered. 
	\item Any malicious HT trying to perform illegal access control or unauthorized data transfer. We consider that the attackers can overwrite the data of a specific data channel and intentionally change the computational output. We also assume that, the malicious attacker could cause potential data leakage by changing the operating mode of the IP core.

	\item Any malicious attacker located in the CPU core, trying to gain unauthorized access or leak sensitive information of other applications.

\end{enumerate}

\noindent
\section {Proposed Architechture }

\begin{figure*}[h!]
	\centerline{\includegraphics[width=13cm]{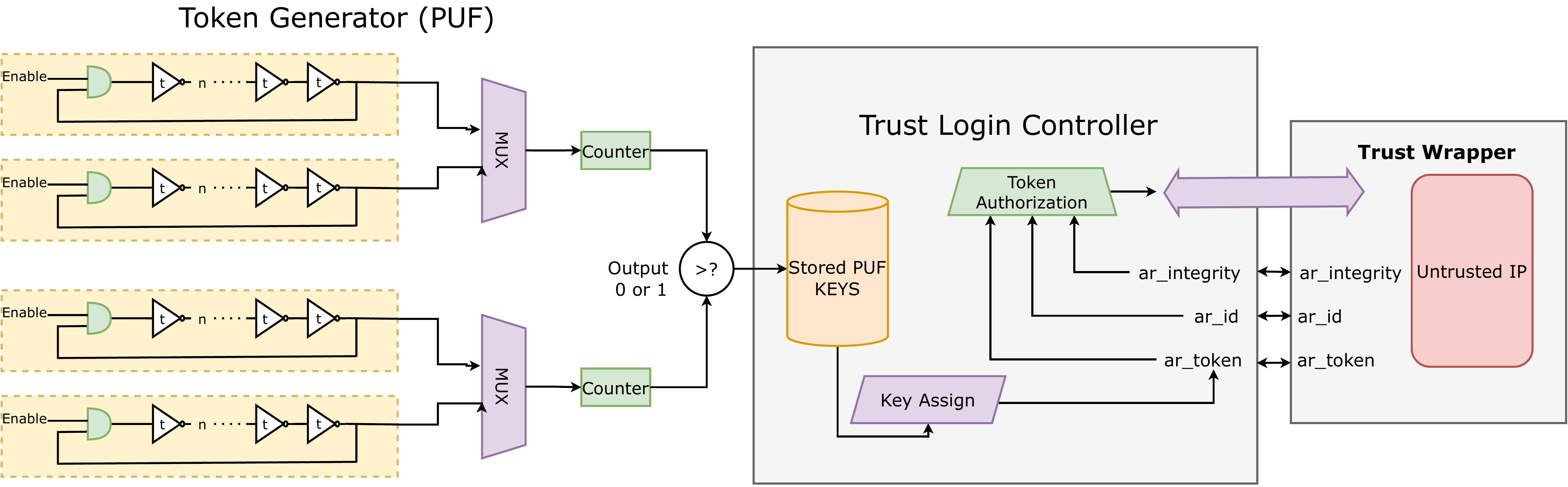}}
	\caption{Overview of the proposed TrustToken architecture framework. Consist of TrustToken Controller, TrustWrapper and TokenGenerator}
	\label{fig:architechture}
	
\end{figure*}

The goal of TrustToken is to provide a root of trust-based runtime isolation enabled mechanism, which allows an SoC owner to provide secure and flexible communication of IP cores without any additional secure storage services or system. 
Figure \ref{fig:architechture} illustrates the detailed architecture of our proposed design, which includes the following components: the TrustToken Controller, TrustWrappers and TokenGenerator.

\paragraph{\textbf{TrustToken Controller.}}

\begin{figure}[h]
	\centerline{\includegraphics[width=7cm]{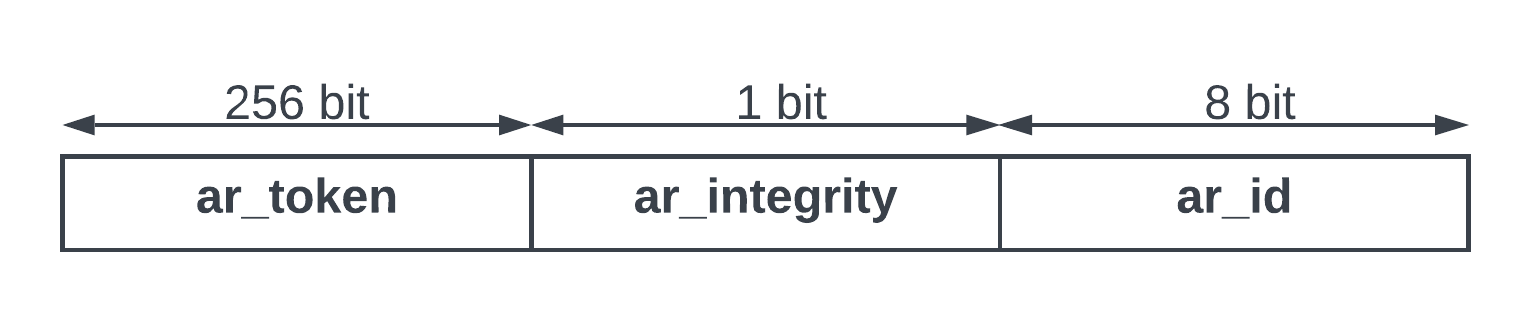}}
	\caption{TrustWrapper data ports: Proposed TrustToken signals with their relative port width.}
	\label{data_bits}
\end{figure}

 The \textbf{TrustToken} controller is a separate centralized IP dedicated to generating unique Tokens/IDs for the IPs and maintaining the security policies in the connected world. Any IP Integrator has to change the token's parameter value named \textbf{\textit{ar\_integrity}} to assert the validity check (Fig  \ref{data_bits}). When this value is assigned as a LOW state, it will disable the isolation feature. When HIGH, it will impose the isolation mechanism of the IP and execute the IP in a non-trusted zone after successful authorization. 
 After generating the keys by PUF module, they are delivered to Central TrustToken Controller for assigning token IDs. Central TrustToken Controller works as the central security headquarters of the whole SoC system and is responsible for distributing all Token IDs provided by the integrated PUF module. This token ID is referred to as \textbf{\textit{ar\_token}} signal, and the length of this signal is 256 bits. Central TrustToken Controller also assigns specific ID for each of the non-trusted IP which is denoted as\textbf{\textit{ ar\_id}}. \textbf{TrustToken} controller randomly distributes the keys among the IP connected to this controller and stores the allocated Token IDs for respective IPs along with Tokens for future verification. Whenever any IP requests a READ/WRITE access to the \textbf{TrustToken} controller, it compares the received Token ID with the securely stored Tokens list. After successful authorization, it will either enable the data channel for communication or restrict it immediately. 
\paragraph{\textbf{Trust Wrapper.}}
\label{sub:isolation}

In our proposed architecture, every IP will be wrapped in a security wrapper labeled as TrustWrapper. TrustWrapper has two different operating interfaces: Secured and Non-secured. Every non-trusted IP core tagged as non-secured will be assigned two additional bus signals to the IP core: ID and Token. Instead of adding any register level isolation mechanism or any separate bus protocol for the secure isolation, we rely on adding extra bus signals to the existing AMBA bus protocol specifications. Adding a separate bus protocol for isolation could create new vulnerabilities and force of modifying the interconnect bridge logic for security check operations.
Further, creating a uniform and unique bus protocol to carry IPs ID and Token information would need a different security mechanism and support every possible bus protocol specifications e.g. bandwidth, channel length, burst size, streaming methods, etc. Each data transaction initiated by the non-trusted IP core will create an authorization request by the Central TrustToken Controller. non-trusted IP should provide valid and unexpired security information (IDs and Token) to the controller block through the security wrapper.

\paragraph{\textbf{Token Generator. }}
\noindent

Due to low overhead and latency, Enhanced Ring Oscillator-based PUF proposed in paper \cite{kawser_puf} is implemented, which is more stable than traditional Ring Oscillator PUF.   Ring Oscillator-based PUF shows promising latency and resource utilization results compared to SRAM PUF, Arbiter PUF, TRNG, or other crypto cores.
 Our custom Ring Oscillator-based PUF solution can generate 256-bit width keys. It has an accepted uniqueness and randomness to fit our goal of providing heterogeneous SoC security. In one of the fundamental research work regarding PUF \cite{kawser_puf}
 strong PUF is defined as the following security properties: 
 1. It will be impossible to clone the PUF circuit physically. 2. It will support many Challenge-Response Pairs(CRPs) so that the adversary cannot mount a brute force attack within a realistic time.
 In terms of the Strong PUF definition, the proposed work can be considered a strong PUF and will be the best candidate to be implemented for the proposed SoC security reason.
 
	

\paragraph{\textbf{Secure Transition between Integrity Level}}
Benhani et al. \cite{benhani_trustzone} showed that any simple malicious TCL script in CAD Tool could comprise the logic state of AWPROT or ARPROT signal in the AXI peripheral or AXI Interconnection in Zynq Based SoC platform. Any malicious TCL script can modify this AWPROT to ARPROT signal to HIGH logic states, which forces the TRUST ZONE controller to execute in a non-secure world even though the IP is insecure, leading to a denial of service attacks. To encounter this problem, we propose a secure transition of Integrity Level Logic. Every transition of the signal \textbf{\textit{(ar\_integrity)}} should go under successful authentication verification by Central \textbf{TrustToken} Controller.

\section{Security Rules}
\label{sec:proposed}
In this section, we describe the formal specifications of the security rules for the \textbf{TrustToken} framework. 
The security formalism defines the security elements and access control primitives that are implemented in the system. Both hardware and software level components are integrated in the security primitives because the software processes offload their to hardware IP cores. The security tuple $\mathbb{S}$ is characterized as follows:
\begin{equation*}
\mathbb{S} := \{U, P, O, T, I, A, D, M\}
\end{equation*} 

\begin{itemize} [leftmargin=*]

	\item $U$ = $\{u_1, u_2, u_3, .... ,u_n\}$ is the set of users in a system.
	\item $P$ = $\{P_1, P_2, P_3, .... ,P_n\}$ is the set of process sets where each user has its corresponding process set $P_i$ = $\{p_{i1}, p_{i2}, p_{i3}, .... ,p_{im}\}$
	\item $O$ = $\{o_1, o_2, o_3, .... ,o_k\}$ is the set of objects. In our proposed framework, objects correspond to various types of non-trusted IP cores.  
	\item $T$ = $\{T_1, T_2, T_3, .... ,T_n\}$ is the set of secret Tokens.
	\item $I$= $\{I_1, I_2, I_3, .... ,I_n\}$ is the set of assigned IDs to each non-trusted IP core.  
		\item $A$ = $\{HIGH,LOW\}$ is the set of integrity access attributes. Here, $HIGH$ is the HIGH state level of integrity, $LOW$ is LOW state level of integrity.
	\item $D$ = $\{yes,no\}$ is the set of decisions. 
	\item $M$ = $\{M_1, M_2, M_3, .... ,M_n\}$ is the set of access matrix. Each user has its corresponding access matrix. Each matrix has $m\times k$ elements where each element is a 3-bit access attribute, $a = a_2a_1a_0$ where $a_2 \rightarrow r, a_1 \rightarrow w, a_0 \rightarrow e$.
\end{itemize}
 
As most of the modern OS system allows us to create multiple user accounts in a single CPU , we include the set of users in the security tuple. Each user can execute  multiple processes and we have included one process under each user.  The integrity access attributes include HIGH and LOW states. To ensure the security of the system, we have defined and established some security rules:

\noindent
\textbf{Rule 1.} For each $u$ $\in$ $U$, there is a function $F_u$$\colon$$P$$\rightarrow$$M$ which must be a one to one function.  
Rule 1 ensures secure isolation of hardware access requests as a process under one user can not gain any unauthorized access of other user. 
\vspace{-0.5mm}

\textbf{Rule 2.} An access request is a 4-tuple $\tau := (u, p, o, t, i, a)$ where $u$ $\in$ $U$, $p$ $\in$ $P_i$, $o$ $\in$ $O$, $t_i$ $\in$ $T$, $i_i$ $\in$ $I$ and $a_i$ $\in$ $A$.

Rule 2 defines the access request where a process under a user account requests for a data transaction from a hardware IP core.\\ 
\noindent
\textbf{Rule 3.} Confidentiality Preserving Rule : If a process $p$ $\in$ $P$ has an integrity attribute, $i$ over an object $o$ $\in$ $O$ and the decision is $d$ $\in$ $D$, the confidentiality is preserved if $a_2$ = $r$ or  $a_0$ = $e$ or both.

\vspace{-0.5mm}
\noindent
\textbf{Rule 4.} Integrity Preserving Rule : If a process $p$ $\in$ $P$ has an access attribute $a$ over an object $o$ $\in$ $O$ and the decision is $d$ $\in$ $D$, the integrity is preserved if $a_1$ = $w$ or  $a_0$ = $e$ or both.

\noindent
\textbf{Rule 5.} The access request of a process $p$ $\in$ $P$ over an object $o$ $\in$ $O$ is granted if the decision is $d$ $\in$ $D$ and $d$ = $yes$.

\noindent
\textbf{Rule 6.} Only the Central Trust Controller or an IP integrator in design phase has the access to modify the access matrix $M_i$ $\in$ $M$.
\vspace{-0.5mm}
\noindent

\vspace{-1.5mm}

\noindent

\section {Protocol Evaluation and case studies  }
In this section, we tried to cover the protocol resiliency in described attack scenarios. 
\subsection{Scenario 1 : Compromising ID signals}
 The Token-based authorization aimed to protect against malicious CAD or RTL modification attacks. As stated above, to achieve hardware modification and gain access control in a non-trusted IP, it was enough to modify only some commands in the script. In our proposed design, the PUF-based Token ID is provided only in run-time conditions and exploits the device's manufacturing variation. Also, as the Token ID is not saved in any memory, we assume that this will protect against any malicious attack of altering the Token ID. In section \ref{sec:threat} we have discussed a possible attack scenario where a software level attack was introduced from an arbitrary application core. The malicious adversaries configure a secured IP core and attempt to gain access to the victim IP by initiating a transaction request from a different IP core. However, Central Trust Controller keeps a record of all assigned IDs and Tokens and their respective source and destination IPs. Since the attacker has made an illegal access request from an outside IP core, this attempt will be compared with the saved credentials and prevented if mismatched.


\subsection{Scenario 2 : Compromising access control } 

In the case of Xilinx TrustZone, \cite{xilinx_trustzone}, at the AXI interconnect level security check is performed, and it plays a critical role in the security. This Interconnect crossbar is also responsible for checking the security status of every transaction on the connected AXI bus, which creates a huge security risk. Any malicious attacker intending to break the security layer can easily control the AXI interconnect crossbar by modifying some security signals. This defect was overcome with the proposed secure design, as the proposed Trust Token Controller has enforced a robust and secure system that makes any access control attack very difficult to take control of the internal signals of Central \textbf{TrustToken} Controller. Central \textbf{TrustToken} Controller is itself encrypted with PUF-based Token ID key and hence restricts any unauthorized access control on this IP.  

\subsection{Scenario 3: Comprising INTEGRITY LEVEL}

Any non-trusted IP connected to the Central \textbf{TrustToken} Controller for secure isolation is determined by the status of INTEGRITY LEVEL signals. As stated before in the thread model section, only an IP Integrator can define the INTEGRITY STATUS in hardware level. Any modification of this signal in runtime conditions will need proper authorization, which gives protection against any hidden CAD or RTL script attack. Also, to alter the status of the protection level, any malicious attacker has to show their PUF-based Token ID of the non-trusted IP. In the work, \cite{benhani_trustzone} benhani et al. showed that only by modifying the Arm TrustZone AWPROT/ARPROT signal any malicious attacker could create a significant Denial of Service (DoS) interruption in the SoC. This scenario can be overcome by the proposed secure transition model, where an alteration request should also go into an additional authorization layer.

\section{Performance Evaluation and Implementation Summary}

This section describes the experimental setup and overhead calculation used for implementing our proposed architecture to evaluate the robustness of the proposed \textbf{TrustToken} framework. The main setup was to efficiently implement the architecture and calculate overhead and latency for data transactions.

\subsection{ Evaluation Infrastructure}
To implement the protocol framework, we have used the Zybo Z7-20 (Xilinx XC7Z020) FPGA board throughout the whole article. This board has two ARM Cortex-A9 processors, clocked at 667 MHz
with 1-GB Memory and Zynq - 7000 FPGA processor. Overhead analysis and performance reports were generated and acquired from Xilinx's Vivado Design Suite platform. 
All experiments reported in this article were performed on
Xilinx XC7Z020 FPGA PL fabric.

\subsection{Protocol Performance}
 We evaluated our proposed TrustToken protocol by implementing and synthesizing on Zybo-Z7-20 board. For evaluation, we have attached TrustWrappers around four symmetric crypto IP cores (AES,DES,TRNG and RSA). Every TrustWrappers was assigned in HIGH integrity state to evaluate the proposed architecture model. We also initiated 5 different applications on ARM processor to access the crypto cores computational results. In our implementation, we successfully introduced trusted execution environment by TrustToken model and observed the results. In section \ref{sec:threat}, we considered a possible software level attack scenario, where a malicious attacker from Application 3 (mapped to TRNG hardware IP core) is trying to establish an authorized access path to RSA IP core. We implemented this scenario, and the attacked was prevented by TrustToken module.  
 To compare the protocol performance, we also designed VIVADO CAD tool based Xilinx TrustZone enclave around the four crypto cores by following the work \cite{xilinx_trustzone_cad} and compared with proposed TrustToken protocol. For Xilinx TrustZone, we successfully launch a simple CAD Tool attack by modifying the \textbf{AWPROT} signal in runtime condition scenario. Similarly, the attack attempt was failed in the proposed method, which explicitly prove the protocol resiliency against CAD tools attack.

\subsection{Token Keys Performance Evaluation}
\begin{figure}[h]
	\centerline{\includegraphics[width=5cm]{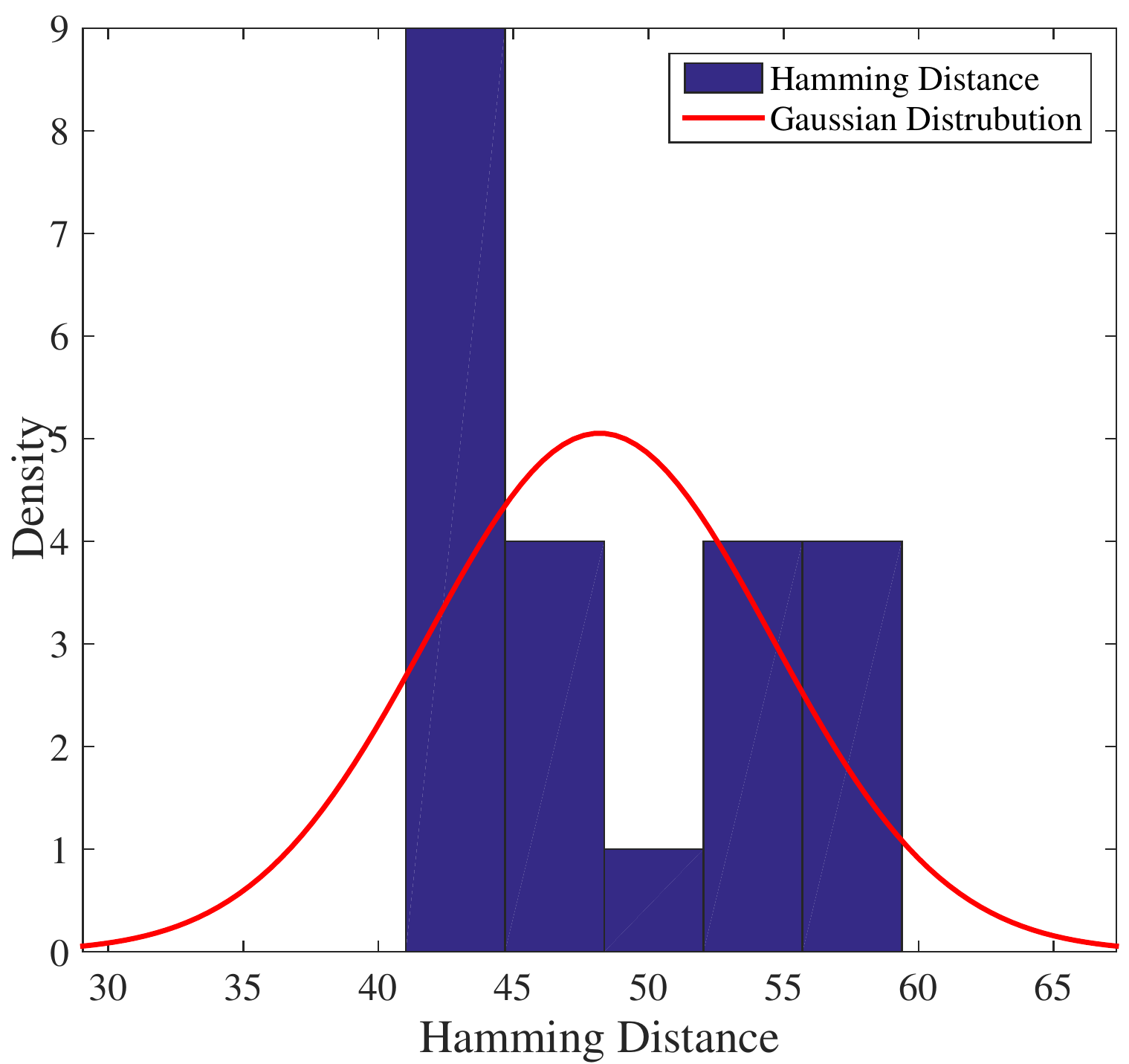}}
	\caption{Hamming Distance between the PUF keys.}
	\label{hamming}
	
\end{figure}

Fig \ref{hamming} shows the hamming distance results calculated from the PUF keys. We can observe from the figure that the hamming distance is closely rounded between 40 and 60 percent, which proves the stability and effectiveness of the keys and is very close to the ideal characteristics of PUF \cite{kawser_puf}.

\begin{table}
	
	\caption{Characterizations of the Ring Oscillator PUF}
	\label{table:PUF}
	\begin{center}
		\begin{tabular}{ |c|c| } 
			\hline
			Parameters  & Values \\ 
			\hline
			\hline
			No of Oscillators & 512 \\
			
			No of Keys Generated & 256  \\
			Length of a single generated Key & 256 bits\\
			Length of a single Challenge Input  & 2 bytes\\
			
			   Randomness & 46.62\% \\
			   Uniqueness & 48.18\%\\
			   Reliability & 100\% \\
			
			\hline
		\end{tabular}
	\end{center}
	
\end{table}
In Table \ref{table:PUF} the overall characterizations of the PUF were summarized. Our internal PUF design includes 512 oscillators and can generates keys of 256 bits wide.

\subsection{Resource  Overhead}
After successful implementation, we have included the utilization report from the VIVADO software platform in Table \ref{table:utlization}. The deployed design shows encouraging results with low resource utilization. BUFG region utilization is only rounded to 6.25 percent. 
\begin{table}[ht]
	\caption{Utilization Report} 
	\centering 
	\begin{tabular}{ |c | c | c |c} 
		\hline 
		Resource  & Available & Utilization (\%) \\ [0.5ex] 
		\hline 
		LUT &   53200 & 618 (1.16\%) \\ 
		
		FF &  106400 & 44 (0.04\%) \\
		BUFG &  32 & 2 (6.25\%) \\[1ex] 
		\hline 
	\end{tabular}
	\label{table:utlization} 
\end{table}

\section{Conclusion }

This paper proposes a Token-based secure SoC architecture for non-trusted IP in a heterogeneous SoC platform. 	\textbf{TrustToken} architecture uses a root of trust-based authorization and provides an extra layer of security against unauthorized access control and attacks. The protocol uses a custom Ring Oscillator-based PUF module to generate keys, and it can exploit the reconfigurable nature of the SoC-based FPGA platform. Our implementation shows low latency and overhead in generating and distributing the PUF keys. We have shown that the proposed protocol uses a constrained LUT and BUFG region of an SoC architecture and effectively provides state-of-the-art illegal software access and data leakage prevention without much resource utilization. This protocol can also be promising for FPGA-based Hardware Accelerators fields etc., Cloud Computing, Machine Learning, and Image Processing. 

\printbibliography

@article{android_tee,
    title = {{Android Enterprise Security White Paper}}
}

@article{pinto_arm_2019,
    title = {{Demystifying Arm TrustZone: A Comprehensive Survey}},
    year = {2019},
    journal = {ACM Comput. Surv},
    author = {Pinto, Sandro},
    volume = {51},
    url = {https://doi.org/10.1145/3291047},
    doi = {10.1145/3291047}
}

@article{xilinx_trustzone_cad,
    title = {{Design a TrustZone-Enalble SoC using the Xilinx VIVADO CAD Tool}}
}

@inproceedings{trojan_1,
    title = {{Hardware Trojan: Threats and emerging solutions}},
    year = {2009},
    booktitle = {2009 IEEE International High Level Design Validation and Test Workshop},
    author = {Chakraborty, Rajat Subhra and Narasimhan, Seetharam and Bhunia, Swarup},
    month = {11},
    publisher = {IEEE},
    doi = {10.1109/hldvt.2009.5340158}
}

@article{tehranipoor_trojan,
    title = {{Introduction to hardware security and trust}},
    year = {2012},
    journal = {Introduction to Hardware Security and Trust},
    author = {Tehranipoor, Mohammad and Wang, Cliff},
    month = {10},
    pages = {1--427},
    volume = {9781441980809},
    publisher = {Springer New York},
    isbn = {9781441980809},
    doi = {10.1007/978-1-4419-8080-9}
}

@article{physicalisolation_huffmire,
    title = {{Moats and drawbridges: An isolation primitive for reconfigurable hardware based systems}},
    year = {2007},
    journal = {Proceedings - IEEE Symposium on Security and Privacy},
    author = {Huffmire, Ted and Brotherton, Brett and Wang, Gang and Sherwood, Timothy and Kastner, Ryan and Levin, Timothy and Nguyen, Thuy and Irvine, Cynthia},
    pages = {281--295},
    isbn = {0769528481},
    doi = {10.1109/SP.2007.28},
    issn = {10816011}
}

@article{op_tee,
    title = {{OP-TEE Documentation TrustedFirmware.org}},
    year = {2021}
}

@article{kawser_puf,
    title = {{Physical Unclonable Function Based Hardware Security for Resource Constraint IoT Devices}},
    year = {2020},
    journal = {IEEE World Forum on Internet of Things, WF-IoT 2020 - Symposium Proceedings},
    author = {Ahmed, Muhammed Kawser and Yanambaka, Venkata P. and Abdelgawad, Ahmed and Yelamarthi, Kumar},
    month = {6},
    publisher = {Institute of Electrical and Electronics Engineers Inc.},
    isbn = {9781728155036},
    doi = {10.1109/WF-IOT48130.2020.9221357}
}

@article{trustzone_white,
    title = {{Programming ARM TrustZone Architecture on the Xilinx Zynq-7000 All Programmable SoC}},
    author = {{Xilinx}},
    url = {https://www.xilinx.com/support/documentation/user_guides/ug1019-zynq-trustzone.pdf}
}

@article{Zhao_trustzone_sram,
    title = {{Providing Root of Trust for ARM TrustZone using On-Chip SRAM}},
    year = {2014},
    author = {Zhao, Shijun and Zhang, Qianying and Hu, Guangyao and Qin, Yu and Feng, Dengguo},
    url = {http://dx.doi.org/10.1145/2666141.2666145.},
    doi = {10.1145/2666141.2666145}
}

@article{Li_attack_1,
    title = {{Research on ARM TrustZone}},
    year = {2019},
    journal = {GetMobile: Mobile Comp. and Comm.},
    author = {Li, Wenhao and Xia, Yubin and Chen, Haibo},
    number = {3},
    month = {1},
    pages = {17--22},
    volume = {22},
    publisher = {Association for Computing Machinery},
    url = {https://doi.org/10.1145/3308755.3308761},
    address = {New York, NY, USA},
    doi = {10.1145/3308755.3308761},
    issn = {2375-0529}
}

@misc{google_sandbox,
    title = {{Sandbox}},
    url = {https://chromium.googlesource.com/chromium/src/+/HEAD/docs/design/sandbox.md}
}

@inproceedings{isolation_7,
    title = {{Secure Hardware Kernels Execution in CPU+FPGA Heterogeneous Cloud}},
    year = {2018},
    booktitle = {2018 International Conference on Field-Programmable Technology (FPT)},
    author = {Hategekimana, Festus and Mandebi Mbongue, Joel and Pantho, Md Jubaer Hossain and Bobda, Christophe},
    pages = {182--189},
    doi = {10.1109/FPT.2018.00035}
}

@article{basak_2017,
    title = {{Security Assurance for System-on-Chip Designs with Untrusted IPs}},
    year = {2017},
    journal = {IEEE Transactions on Information Forensics and Security},
    author = {Basak, Abhishek and Bhunia, Swarup and Tkacik, Thomas and Ray, Sandip},
    number = {7},
    month = {7},
    pages = {1515--1528},
    volume = {12},
    publisher = {Institute of Electrical and Electronics Engineers Inc.},
    doi = {10.1109/TIFS.2017.2658544},
    issn = {15566013}
}

@inproceedings{ray_policy_2015,
    title = {{Security policy enforcement in modern SoC designs}},
    year = {2015},
    booktitle = {2015 IEEE/ACM International Conference on Computer-Aided Design (ICCAD)},
    author = {Ray, Sandip and Jin, Yier},
    pages = {345--350},
    doi = {10.1109/ICCAD.2015.7372590}
}

@inproceedings{isolation_1,
    title = {{Shielding non-trusted IPs in SoCs}},
    year = {2017},
    booktitle = {2017 27th International Conference on Field Programmable Logic and Applications (FPL)},
    author = {Hategekimana, Festus and Whitaker, Taylor and Pantho, Md Jubaer Hossain and Bobda, Christophe},
    pages = {1--4},
    doi = {10.23919/FPL.2017.8056848}
}

@misc{xilinx_trustzone,
    title = {{startup • Vitis Unified Software Platform Documentation: Embedded Software Development (UG1400) • Reader • Documentation Portal}},
    url = {https://docs.xilinx.com/r/en-US/ug1400-vitis-embedded/startup}
}

@article{trojan_2,
    title = {{Ten years of hardware Trojans: a survey from the attacker's perspective}},
    year = {2020},
    journal = {IET Computers {\&} Digital Techniques},
    author = {Xue, Mingfu and Gu, Chongyan and Liu, Weiqiang and Yu, Shichao and O'Neill, Máire},
    number = {6},
    month = {11},
    pages = {231--246},
    volume = {14},
    publisher = {The Institution of Engineering and Technology},
    url = {https://onlinelibrary.wiley.com/doi/full/10.1049/iet-cdt.2020.0041 https://onlinelibrary.wiley.com/doi/abs/10.1049/iet-cdt.2020.0041 https://ietresearch.onlinelibrary.wiley.com/doi/10.1049/iet-cdt.2020.0041},
    doi = {10.1049/IET-CDT.2020.0041},
    issn = {1751-861X}
}

@article{benhani_trustzone,
    title = {{The Security of ARM TrustZone in a FPGA-Based SoC}},
    year = {2019},
    journal = {IEEE Transactions on Computers},
    author = {Benhani, E M and Bossuet, L and Aubert, A},
    number = {8},
    pages = {1238--1248},
    volume = {68},
    doi = {10.1109/TC.2019.2900235}
}

@article{Zhao_trustzone_token,
    title = {{TrustTokenF: A generic security framework for mobile two-factor authentication using TrustZone}},
    year = {2015},
    journal = {Proceedings - 14th IEEE International Conference on Trust, Security and Privacy in Computing and Communications, TrustCom 2015},
    author = {Zhang, Yingjun and Zhao, Shijun and Qin, Yu and Yang, Bo and Feng, Dengguo},
    month = {12},
    pages = {41--48},
    volume = {1},
    publisher = {Institute of Electrical and Electronics Engineers Inc.},
    isbn = {9781467379519},
    doi = {10.1109/TRUSTCOM.2015.355}
}

@article{samsung_knox,
    title = {{White Paper: An Overview of the Samsung KNOX TM Platform}},
    year = {2016}
}

\end{document}